\documentclass[a4paper, 12pt]{article}

\usepackage{amsmath, amssymb, amsthm}
\usepackage[english]{babel}
\usepackage[T1]{fontenc}

\newcommand{\be}{\begin{equation}}
\newcommand{\ee}{\end{equation}}
\newcommand{\bd}{\begin{displaymath}}
\newcommand{\ed}{\end{displaymath}}
\newcommand{\ba}{\begin{eqnarray}}
\newcommand{\ea}{\end{eqnarray}}

\def\R{{I \!\! R}}

\def\v12{(v-w)}

\def\({\left(}
\def\){\right)}

\def\bgr#1\egr{{\allowdisplaybreaks\begin{gather}#1\end{gather}}}
\def\bma#1\ema{{\allowdisplaybreaks\begin{align}#1\end{align}}}

\def\oplem#1{\begin{lemma}\, {\rm #1}\, \it }
\def\cllem{\end{lemma}\rm \par }
\def\opthm#1{\begin{theorem}\, {\rm #1}\, \it }
\def\clthm{\end{theorem}\rm \par }

\def\R{\mathbb{R}}

\newcommand{\fer}[1]{(\ref{#1})}
\newcommand{\bq}{\begin{equation}}
\newcommand{\eq}{\end{equation}}
\def\bqa{\begin{eqnarray}}
\def\eqa{\end{eqnarray}}
\def\bd{\begin{displaymath}}
\def\ed{\end{displaymath}}

\newtheorem{thm}{Theorem}
\newtheorem{cor}[thm]{Corollary}

\theoremstyle{remark}
\newtheorem{rem}[thm]{Remark}
\theoremstyle{definition}

\newenvironment{equations}{\equation\aligned}{\endaligned\endequation}

\begin{document}

\title{A strengthened entropy power inequality for log-concave densities}

 \author{Giuseppe Toscani \thanks{Department of Mathematics, University of Pavia, via Ferrata 1, 27100 Pavia, Italy.
\texttt{giuseppe.toscani@unipv.it} }}

\maketitle

\begin{center}\small
\parbox{0.85\textwidth}{
\textbf{Abstract.} We  show that Shannon's entropy--power inequality admits a
strengthened version in the case in which the densities are log-concave. In such a
case, in fact, one can extend the Blachman--Stam argument \cite{Sta, Bla} to obtain a
sharp inequality for the  second derivative of Shannon's entropy functional with
respect to the heat semigroup.

\medskip
\textbf{Keywords.} Entropy-power inequality, Blachman--Stam inequality.}
\end{center}

\medskip

\section{Introduction}

Given a random vector $X$ in $\R^n$, $n \ge 1$, with a smooth, rapidly decaying
probability density $f(x)$ such that $\log f$ has growth at most polynomial at
infinity, consider the functional
 \begin{equations}\label{j2}
 J(X)= J(f) = \sum_{i,j=1}^n \int_{\{f>0\}} \left[\partial_{ij}(\log f) \right]^2f \, dx = \\
\sum_{i,j=1}^n \int_{\{ f>0\}} \left[\frac{\partial_{ij}f}f -
\frac{\partial_{i}f\partial_{j}f}{f^2} \right]^2 f \, dx.
 \end{equations}
This functional is closely related to  the entropy functional (or Shannon's entropy)
of the random vector $X$
 \be\label{Shan}
H(X) = H(f) = - \int_{\R^n} f(x) \log f(x)\, dx.
 \ee
through the solution $u(x,t)$ of the initial value problem for the heat equation in
the whole space $\R^n$,
 \be\label{heat0}
 \frac{\partial u}{\partial t} = \kappa \Delta u, \qquad u(x,t=0) = f(x),
 \ee
where $\kappa$ is the diffusion constant. To our knowledge, in one-dimension this link
was first remarked in 1965 by McKean  \cite{McK}, who computed the evolution in time
of the subsequent derivatives of the entropy functional $H(u(t))$ of the solution to
\fer{heat0}, with the diffusion constant $\kappa=1$.  McKean computations at the first
two orders gave the identities
 \be\label{der1}
I(f)= \left. \frac d{dt}\right|_{t=0} H(u(t)); \qquad \left. J(f)= -\frac 12 \frac
d{dt}\right|_{t=0} I(u(t)) .
 \ee
In \fer{der1}  $I(f)$  defines, for a given  random vector $X$ in $\R^n$,  its Fisher
information
 \be\label{fish}
I(X) = I(f) = \int_{\{f>0\}} \frac{|\nabla f(x)|^2}{f(x)} \, dx.
 \ee
Hence, $J(u(t))$ can be interpreted as a measure of the (nonnegative) \emph{entropy
production} of the Fisher information when evaluated along the solution to the heat
equation.

While the importance of the first identity in \fer{der1}, well-known in information
theory with the name of  DeBruijn's identity is well established \cite{CoT}, the role of the
second identity in \fer{der1} seems to be restricted to its use in the proof of the
concavity of entropy power \cite{Cos}, and to the so-called entropy-entropy methods
\cite{AMTU, BEm}, where it has been shown highly useful in the proof of the
logarithmic Sobolev inequality \cite{Tos}.

In this paper we study inequalities related to the functional $J(\cdot)$, when evaluated on
convolutions.
 The main result  is a new inequality for $J(X+Y)$, where
$X$ and $Y$ are independent random vectors in $\R^n$, such that their probability
densities $f$ and $g$ are \emph{log-concave}, and $J(X)$, $J(Y)$ are well defined. By
resorting to an argument close to that used by Blachman \cite{Bla} in his original
proof of entropy power inequality, for any given pair of positive constants $a,b$, we
prove the bound
 \begin{equation}\label{ine-main}
 J(f*g) \le \frac{a^4}{(a+b)^4} J(f) + \frac{b^4}{(a+b)^4}J(g) + \frac{2a^2b^2}{(a+b)^4} H(f,g),
 \end{equation}
where
 \be\label{rest}
 H(f,g) = \sum_{i,j=1}^n \int_{\{f>0\}} \frac{\partial_i f\partial_j f}f \, dx
 \int_{\{g>0\}} \frac{\partial_i g\partial_j g}g \, dx.
 \ee
Note that, in one-dimension $H(f,g)$ coincides with the product of the Fisher
information of $f$ and $g$, $H(f,g) = I(f)I(g)$. Inequality \fer{ine-main} is sharp.
Indeed, there is equality if and only if $X$ and $Y$ are $n$-dimensional Gaussian
vectors with covariance matrices proportional to $aI_n$ and $bI_n$ respectively, where
$I_n$ is the identity matrix.

Even if inequality \fer{ine-main} is restricted to the set of log-concave densities,
this set  includes many of the most commonly-encountered parametric families of
probability density functions \cite{MO}. Among other properties, log-concave densities are
stable under convolution. If $f$ and $g$ are (possibly multidimensional) log-concave
densities, then their convolution $f*g$ is log-concave.

Optimizing over $a$ and $b$, one obtains from \fer{ine-main} the inequality
  \be\label{sharp2}
\frac 1{\sqrt{J(X+Y)}} \ge \frac 1{\sqrt{J(X)}}+ \frac 1{\sqrt{J(Y)}},
 \ee
where, also in this case, equality holds if and only if both $X$ and $Y$ are Gaussian
random vectors with proportional covariance matrices.

Inequality \fer{sharp2} shows that, at least if applied to log-concave probability
densities, the functional $J(\cdot)$ behaves with respect to convolutions like
Shannon's entropy power \cite{Sha, Sta} and Fisher information \cite{Sta, Bla}.
Actually, Shannon's entropy power inequality, due to Shannon and Stam \cite{Sha, Sta}
(cf. also \cite{Cos, GSV, GSV2, Rio, ZF} for other proofs and extensions) gives a lower bound on Shannon's entropy power of the sum of independent random
variables $X, Y$ with values in $\R^n$ with densities
 \be\label{entr}
N(X+Y) \ge N(X) + N(Y),
 \ee
with equality if and only $X$ and $Y$ are Gaussian random vectors with proportional
covariance matrices. In \fer{entr} the entropy-power of the random variable $X$ with
values in $\R^n$ is defined by
 \be\label{ep}
 N(X) = N(f) = \exp\left(\frac 2n H(X)\right).
 \ee
With no doubts, this is one of the fundamental information theoretic inequalities
\cite{DCT}.

Likewise,  Blachman--Stam inequality \cite{Sta, Bla} gives a lower
 bound on the inverse of Fisher information of the sum of independent random vectors
with (smooth) densities
 \be\label{BS}
\frac 1{I(X+Y)} \ge \frac 1{I(X)}+ \frac 1{I(Y)},
 \ee
still with equality if and only $X$ and $Y$ are Gaussian random vectors with
proportional covariance matrices.

The fact that inequalities \fer{sharp2}, \fer{entr} and \fer{BS} share a common nature
is clarified by noticing that, when evaluated in correspondence to a Gaussian vector
$X$ with covariance matrix $\sigma I_n$, the three (related by \fer{der1}) functionals
$N(X)$, $1/I(X)$ and $1/\sqrt{J(X)}$ are linear functions of $\sigma$.

An interesting application of inequality \fer{ine-main} is linked to the evolution in
time of the functional
 \be\label{lh}
 \Lambda(t)= \Lambda(f(t),g(t)) = H(f(t)*g(t)) - \kappa H(f(t)) -(1-\kappa)H(g(t)),
 \ee
where, for a positive constant $\kappa$, with $0<\kappa<1$, $f(x,t)$ (respectively
$g(x,t)$) are the solutions to the heat equation \fer{heat0} with diffusion constant
$\kappa$ (respectively $1-\kappa$), corresponding to the initial data $f$ and $g$ which are log-concave
probability densities in $\R^n$. Since, for $\alpha
>0$
 \[
 H(f_\alpha) = H(f) - \log \alpha,
 \]
the functional $\Lambda(t)$ is dilation invariant, that is invariant with respect to the
scaling
 \be\label{scal}
 f(x) \to f_\alpha(x) = {\alpha^n} f\left( {\alpha} x \right), \quad \alpha >0.
 \ee
In \cite{Tos3} we proved that $\Lambda(t)$ is monotonically decreasing in time from $\Lambda
(0)$ to
 \[
\lim_{t\to \infty} \Lambda(t) = -\frac n2 \left[ \kappa \log \kappa +
(1-\kappa)\log(1-\kappa)\right],
 \]
which implies the inequality
 \[
H(f*g) - \kappa H(f) -(1-\kappa)H(g) \ge -\frac n2 \left[ \kappa \log \kappa +
(1-\kappa)\log(1-\kappa)\right].
 \]
By optimizing over $\kappa$, this inequality implies the entropy power inequality
\fer{entr}. By choosing now log-concave densities $f$ and $g$ as initial data in the
heat equations, we can prove, in consequence of inequality \fer{ine-main}, that the
functional $\Lambda(t)$ is a convex function of time, and this implies, optimizing
over $\kappa$, the strengthened entropy power inequality
 \be\label{sep}
N(X+Y) \ge \left[N(X) + N(Y)\right]R(X,Y),
 \ee
where the quantity $R(X,Y)\ge 1$ can be interpreted as a measure of the
\emph{non-Gaussianity} of the two random vectors $X,Y$. Indeed, $R(X,Y) = 1$ if and
only if both $X$ and $Y$ are Gaussian random vectors.

Inequality \fer{lh} describes a new property of convolutions of Gaussian densities.
Among all solutions to the heat equation with log-concave densities as initial data,
the Gaussian self-similar solutions are the unique ones for which the functional
$\Lambda(t)$ remains constant in time. In all the other cases $\Lambda(t)$ is a
\emph{convex} function of time. This property is reminiscent of the well-known
\emph{concavity of entropy power} theorem, which asserts that, given the solution
$u(x,t)$ to the heat equation \fer{heat0},
 \be\label{conc}
\frac{d^2}{dt^2}\left(N(u(t))\right) \le 0.
 \ee
Inequality \fer{conc} is due to Costa \cite{Cos}. Later, the proof has been simplified
in \cite{Dem, DCT}, by an argument based on the Blachman--Stam inequality \cite{Bla}.
Moreover, a short and simple proof has been obtained by Villani \cite{Vil}, using
ideas from the aforementioned McKean paper \cite{McK}. It is interesting to notice
that  the concavity property has been extended to Renyi entropy power, when evaluated
along the solution to a nonlinear diffusion equation \cite{ST}.

These studies reveal that the heat equation (and, more in general, the nonlinear
diffusion equations) are  quite useful instruments to be used in connection with the
proof of inequalities. In some recent papers \cite{Tos1, Tos2, Tos3}, this connection
has been made evident by showing that, in addition to Shannon's entropy power
inequality,  a number of inequalities in sharp form, like the classical Young's
inequality and its converse \cite{BL}, the Brascamp--Lieb type inequalities \cite{BL},
Babenko's inequality \cite{Bab}, Pr\'ekopa--Leindler inequality \cite{Pr1}, Nash's
inequality and the logarithmic Sobolev inequality follow by monotonicity arguments of
various Lyapunov functionals, when evaluated on solutions to the heat equation.

A careful reading of Dembo's proof of the concavity of entropy power \cite{Dem}
clarifies once more the connections among the functionals $H,I$ and $J$. Actually,
Dembo proved inequality \fer{conc} in the equivalent form
 \be\label{not}
 J(f) \ge \frac 1n I(f)^2, \qquad J(f) = -\frac 12 \left. \frac d{dt}\right|_{t=0}
 I(f*M_{2t}),
 \ee
that is reminiscent of \fer{der1}.

As already mentioned in this introduction, while the functional $J(f)$ has been
introduced in Shannon's theory in connection with the proof of the concavity of
Shannon's entropy power, so that it appears in the related literature after Costa's
paper \cite{Cos} in $1985$, in one dimension of space, various properties of the
$J(f)$ functional were considered by McKean in its pioneering paper on Kac's
caricature of Maxwell molecules in kinetic theory of rarefied gases \cite{McK} in
$1965$. In his paper, McKean investigated various connections between Fisher
information \fer{fish} and its derivative $J(f)$ along the heat flux, motivated by
proving the old conjecture that subsequent derivatives of Shannon's entropy along the
solution to the heat equation alternate in sign. McKean original inequalities for the
functional $J(f)$ were subsequently generalized to higher dimensions to give a new
proof of the logarithmic Sobolev inequality with an explicit remainder \cite{Tos}.

In more details, Section \ref{sec2} will be devoted to the proof of inequality
\fer{ine-main}. For the sake of clarity, we will present first the proof in one
dimension of space. Then, a multi-dimensional version will be obtained resorting to
some additional result. In Section \ref{secEP} we will show how inequality
\fer{ine-main} could be fruitfully used to obtain, for log-concave densities, the
strengthened version \fer{sep} of the entropy power inequality. Unfortunately, it
seems quite difficult to express the additional term $R(X,Y)$ in \fer{sep} in the form
of a \emph{distance} of the involved densities from the Gaussian density, and we live
we leave this question to future research.

\section{A new inequality for convolutions}\label{sec2}

 \subsection{Log-concave functions and scores}

 We recall that a function $f$ on $\R^n$ is log-concave if it is of the
form
\be\label{log-c}
 f(x) = \exp\left\{-\Phi(x)\right\},
 \ee
for some convex function $\Phi: \R^n \to \R$. A prime example is the Gaussian density,
where $\Phi(x)$ is quadratic in $x$. Further, log-concave distributions include Gamma
distributions with shape parameter at least one, $Beta(\alpha, \beta)$ distributions
with $\alpha, \beta \ge1$, Weibull distributions with shape parameter at least one,
Gumbel, logistic and Laplace densities (see, for example, Marshall and Olkin
\cite{MO}). Log-concave functions have a number of properties that are desirable for
modelling.  Marginal distributions, convolutions and product measures of log-concave
distributions and densities are again log-concave (cf. for example, Dharmadhikari and
Joag-Dev \cite{DJ}).

A main consequence of log-concavity, which is at the basis of most computations in this paper, is the following. Consider the heat
equation \fer{heat0} in $\R^n$, $n \ge 1$.
If $M_\sigma(x)$ denotes the Gaussian density in $\R^n$ with zero mean and covariance matrix $\sigma I_n$
 \be\label{max}
M_\sigma(x) = \frac 1{(2\pi \sigma)^{n/2}}\exp\left(- \frac{|x|^2}{2\sigma}\right),
 \ee
 the solution at time $t$ to the heat equation \fer{heat0} coincides with  $u= f*M_{2\kappa t}$.

Assume that the initial datum $f(x)$ is a non-negative, log-concave integrable
function. Then, at each subsequent time $t>0$, the solution $u(\cdot,t)$ to the heat
equation, convolution of the log-concave functions $f$ and the Gaussian density
$M_{2\kappa t}$ defined in \fer{max}, is a non-negative integrable log-concave
function.  In other words, the heat equation propagates log-concavity.

This simple remark, allows to proof things by using smooth functions with fast decay
at infinity.

It is interesting to notice that the expressions of Shannon's entropy $H$, Fisher
information $I$ and Fisher's entropy production $J$ take a very simple form if
evaluated in correspondence to log-concave densities $f$, when written as in
\fer{log-c}. In this case, if $X$ is a random vector in $\R^n$ with density $f$, these
functionals can be easily recognized as moments of $\Phi(X)$ or of its derivatives. It
is immediate to reckon that Shannon's entropy $H$ coincides with
 \be\label{hl}
 H(f) = \int_{\R^n} \Phi(x) f(x) \, dx.
 \ee
 The Fisher information $I$ reads
  \be\label{il}
  I(f) = \int_{\R^n} |\nabla \Phi(x)|^2 f(x) \, dx = \sum_{i=1}^n\int_{\R^n} |\partial_i\Phi(x)|^2 f(x) \, dx,
  \ee
 and, last, Fisher's entropy production $J$ takes the form
  \be\label{pl}
  J(f) =  \sum_{i,j=1}^n\int_{\R^n} |\partial_{ij}\Phi(x)|^2 f(x) \, dx.
  \ee
Thus, the functionals are well-defined in terms of the convex function $\Phi$
characterizing the log-concave function $f$.

  For the log-concave Gaussian density \fer{max}
   \[
   \Phi(x) = \frac{|x|^2}{2\sigma} +\frac n2 \log 2\pi\sigma,
   \]
 which implies, for $i,j= 1,2, \dots , n$
  \[
  \partial_i \Phi(x) = \frac{x_i}\sigma, \quad \partial_{ij}\Phi(x) = \frac 1\sigma \delta_{ij},
  \]
  where, as usual, $\delta_{ij}$ is the Kronecker delta.

According to the standard definition, given a random vector $X$ in $\R^n$ distributed
with with absolutely continuous probability density function $f(x)$
 \be\label{sco1}
\rho(X) = \frac{\nabla f(X)}{f(X)} ,
 \ee
denotes the (almost everywhere defined) score function of the random variable
\cite{CH} (cf. also \cite{BM} for further details). The score has zero mean, and its
variance is just the Fisher information. For log-concave densities, which are
expressed in the form \fer{log-c}
 \be\label{sc1}
 \rho(X) = -\nabla \Phi(X)
 \ee
In view of definition \fer{j2} and \fer{sco1} one can think to introduce the concept
of second-order score of a random vector $X$ in $\R^n$, defined by the symmetric
Hessian matrix $\mathcal{H}(X)$ of $-\log f(X)$, with elements
 \be\label{scor2}
 \Psi_{ij}(X) =
\frac{\partial_{i}f\partial_{j}f(X)}{f^2(X)} -\frac{\partial_{ij}f(X)}{f(X)} .
 \ee
Then, as the Fisher information coincides the second moment of the score function, the
functional $J(X)$ in \fer{j2} is expressed by the moment of the trace of the product
matrix $\mathcal{H}(X)\cdot\mathcal{H}(X)$.  For a log-concave function, the element
$\Psi_{i,j}$ of the Hessian matrix $\mathcal{H}(X)$ defining the second-order score
function takes the simple expression
 \be\label{sc2}
\Psi_{ij}(X) = \partial_{ij}\Phi(X).
 \ee
Note that a Gaussian vector $M_\sigma$ is uniquely defined by a \emph{linear} score function $\rho(M_\sigma) = M_\sigma/\sigma$ and by a \emph{constant} second-order score matrix $ \mathcal{H}(M_\sigma) = I_n/\sigma$.

\subsection{The one-dimensional case}

For the moment, let us fix $n=1$. In the rest of this section, we will only consider smooth log-concave
probability densities $f(x)$ (cf. definition \fer{log-c}) such that $\Phi(x) =-\log f(x)$  has growth at most polynomial at infinity. In order not to worry about derivatives of
logarithms, which will often appear in the proof, we may also impose that $|\Phi'(x)/\Phi(x)| \le C(1 + |x|^2)$ for some positive constant $C$. The general case will easily follow by a density argument \cite{LT}.
 Let
\[
k(x) = f*g(x).
\]
The main argument here is due to Blachman \cite{Bla}, who proved in this way
inequality \fer{BS}. Since for any pair of  positive constants  $a,b$ we have the
identity
 \[
(a+b)k'(x) = a \int_\R f'(x-y)g(y) \, dy + b \int_\R f(x-y)g'(y) \, dy,
 \]
dividing by $k(x)>0$ we obtain
 \[
(a+b)\frac{k'(x)}{k(x)}  = a \int_\R \frac{f'(x-y)}{f(x-y)}
\frac{f(x-y)g(y)}{k(x)} \, dy + b \int_\R
\frac{g'(y)}{g(y)}\frac{f(x-y)g(y)}{k(x)} \, dy =
 \]
 \[
\int_\R \left(a \frac{f'(x-y)}{f(x-y)} + b \frac{g'(y)}{g(y)}\right) \,
d\mu_x(y).
 \]
We denoted
 \[
d\mu_x(y) = \frac{f(x-y)g(y)}{k(x)} \, dy.
 \]
Note that, for every $x \in \R$, $d\mu_x$ is a unit measure on $\R$.
Consequently, by Jensen's inequality
 \[
(a+b)^2 \left(\frac{k'(x)}{k(x)}\right)^2 = \left[ \int_\R \left(a
\frac{f'(x-y)}{f(x-y)} + b \frac{g'(y)}{g(y)}\right) \, d\mu_x(y)
\right]^2 \le
 \]
 \be\label{jen}
\int_\R \left(a \frac{f'(x-y)}{f(x-y)} + b \frac{g'(y)}{g(y)}\right)^2 \,
d\mu_x(y).
 \ee
On the other hand, by analogous argument, for any pair of  positive constants  $a,b$ we have the identity
 \[
(a+b)^2 k''(x) = a^2 \int_\R f''(x-y)g(y) \, dy + b^2 \int_\R f(x-y)g''(y) \, dy +2ab\int_\R f'(x-y)g'(y) \, dy.
 \]
Thus, dividing again by $k(x)>0$ we obtain
 \be\label{d2}
(a+b)^2\frac{k''(x)}{k(x)}  = \int_\R \left( a^2 \frac{f''(x-y)}{f(x-y)} +
 b^2 \frac{g''(y)}{g(y)} +2ab \frac{f'(x-y)}{f(x-y)}\frac{g'(y)}{g(y)}\right)  \,
d\mu_x(y).
 \ee
If we subtract identity \fer{d2} from inequality \fer{jen} we conclude with the inequality
 \begin{equations}\label{ok1}
(a+b)^2 \left[ \left(\frac{k'(x)}{k(x)}\right)^2 - \frac{k''(x)}{k(x)} \right] \le \qquad\qquad\qquad\qquad\qquad\\
\int_\R \left\{ a^2 \left[ \left(\frac{f'(x-y)}{f(x-y)}\right)^2 - \frac{f''(x-y)}{f(x-y)}\right] + b^2 \left[ \left(\frac{g'(y)}{g(y)}\right)^2 - \frac{g''(y)}{g(y)}\right]\right\} \,
d\mu_x(y).
\end{equations}
It is important to note that, since the functions $f,g$ (and consequently $k$) are log-concave, the left-hand side of inequality \fer{ok1} is non-negative. Therefore, taking the square on both sides of \fer{ok1}, and using once more Jensen's inequality we end up with the inequality
 \begin{equations}\label{jen2}
(a+b)^4 \left[ \left(\frac{k'(x)}{k(x)}\right)^2 - \frac{k''(x)}{k(x)} \right]^2 \le \qquad\qquad\qquad\qquad\qquad\\
\int_\R \left\{ a^2 \left[ \left(\frac{f'(x-y)}{f(x-y)}\right)^2 - \frac{f''(x-y)}{f(x-y)}\right] + b^2 \left[ \left(\frac{g'(y)}{g(y)}\right)^2 - \frac{g''(y)}{g(y)}\right]\right\}^2 \,
d\mu_x(y).
\end{equations}
Multiplying both sides of \fer{jen2} by $k(x)$, and integrating over $\R$ yields the inequality
 \[
 (a+b)^4 J(k) \le  a^4 J(f) + b^4 J(g) + 2a^2 b^2 I(f) I(g).
 \]
Indeed, in one dimension, definition \fer{j2} of the functional $J(\cdot)$ reduces to
 \be\label{j2-1}
 J(f) = \int_{\{f>0\}}  \left[ \left(\frac{f'(x)}{f(x)}\right)^2 - \frac{f''(x)}{f(x)} \right]^2 f(x)\, dx.
 \ee
Moreover
 \[
\int_\R dx \int_\R dy \left[ \left(\frac{f'(x-y)}{f(x-y)}\right)^2 - \frac{f''(x-y)}{f(x-y)}\right] \left[ \left(\frac{g'(y)}{g(y)}\right)^2 - \frac{g''(y)}{g(y)}\right] f(x-y)g(y) =
 \]
 \[
\int_\R dx \int_\R dy  \left(\frac{f'(x-y)}{f(x-y)}\right)^2 \left(\frac{g'(y)}{g(y)}\right)^2 f(x-y)g(y) = I(f)I(g),
 \]
where $I(f)$ (respectively $I(g)$) denotes the Fisher information  of $f$
(respectively $g$)
 \[
 I(f) = \int_{\{ f>0 \}} \frac{(f'(x))^2}{f(x)} \, dx.
 \]
 The cases of equality in \fer{jen} and \fer{jen2} are easily
found resorting to the following argument. Equality follows if, after
application of Jensen's inequality,  there is equality in \fer{jen}. On
the other hand, for any convex function $\varphi$ and unit measure $d\mu$
on the set $\Omega$, equality in Jensen's inequality
 \[
 \varphi(\int_\Omega f\,d\mu) \le \int_\Omega \varphi(f) \, d\mu
 \]
holds true if and only if $f$ is constant, so that
 \[
f = \int_\Omega f\,d\mu.
 \]
In our case, this means that there is equality if and only if the function
 \[
a \frac{f'(x-y)}{f(x-y)} + b \frac{g'(y)}{g(y)}
 \]
does not depend on $y$. If this is the case, taking the derivative with
respect to $y$, and using the identity
 \[
\frac d{dy}\left(\frac{f'(x-y)}{f(x-y)}\right) = -\frac
d{dx}\left(\frac{f'(x-y)}{f(x-y)}\right),
 \]
we conclude that $f$ and $g$ are forced to satisfy
 \be\label{eq3}
a \frac{d^2}{dx^2}\log f(x-y) = b \frac{d^2}{dy^2}\log g(y).
 \ee
Note that \fer{eq3} can be verified if and only if the functions on both
sides are constant. Thus, there is equality if and only if
 \be\label{m2}
\log f(x) = b_1x^2 + c_1 x + d_1, \quad \log g(x) = b_2x^2 + c_2 x + d_2 .
 \ee
By coupling \fer{m2} with \fer{eq3}, we obtain that there is equality in \fer{fish} if
and only if $f$ and $g$ are Gaussian densities, of variances $ca$ and $cb$,
respectively, for any given positive constant $c$. Analogous argument leads to the
same conclusion as far as inequality \fer{jen2} is concerned. We proved

\begin{thm}\label{bl} Let $f(x)$ and $g(x)$  be log-concave probability density
functions with values in $\R$,  such that both $J(f)$ and $J(g)$, as given by
\fer{j2-1} are bounded. Then, also $J(f*g)$ is bounded, and for any pair of positive
constants $a,b$
 \be\label{fin1}
  J(f*g) \le  \frac{a^4}{(a+b)^4} J(f) + \frac{b^4}{(a+b)^4} J(g) + 2\frac{a^2 b^2}{(a+b)^4} I(f) I(g).
 \ee
Moreover, there is equality in \fer{fin1} if and only if, up to translation and
dilation  $f$ and $g$ are Gaussian densities, $f(x)= M_{a}(x)$ and $g(x) = M_{b}(x)$.
\end{thm}

\begin{rem}
The condition of log-concavity enters into the proof of Lemma \ref{bl} when we pass from inequality \fer{ok1} to inequality \fer{jen2}. Without the condition of log-concavity, in fact, the left-hand side of \fer{ok1} has no sign, and \fer{jen2} does not hold true. Of course, this fact does not exclude the possibility that  inequality \fer{fin1} could hold also for other classes of probability densities, but if any, another method of proof has to be found, or a counterexample is needed.
\end{rem}

Theorem \ref{bl} allows to prove inequality \fer{sharp2}. To this aim, note that, for any pair of positive constants $a,b$
 \[
 2\sqrt{J(f)J(g)} \le \frac ab J(f) + \frac ba J(g).
 \]
Moreover, as proven first by Dembo \cite{Dem},  and later on by Villani \cite{Vil} with a proof based on McKean ideas \cite{McK} \fer{not} implies
 \be\label{d3}
 J(f) \ge I(f)^2.
 \ee

\begin{rem}
The proof of \fer{d3} is immediate and enlightening.  Given the random variable $X$ distributed with a sufficiently smooth density $f(x)$, consider the (almost everywhere defined) second-order score variable (cf. definition \fer{scor2})
 \be\label{sco2}
 \Psi(X)  =  \left(\frac{f'(X)}{f(X)}\right)^2 - \frac{f''(X)}{f(X)} .
 \ee
Then, denoting with $\langle Y\rangle$ the mathematical expectation of the random variable $Y$, it holds
 \[
 I(f)= I(X) = \langle \Psi(X)\rangle, \qquad J(f) = J(X) = \langle \Psi(X)^2\rangle.
 \]
Then, \fer{d3} coincides with the standard inequality $\langle \Psi(X)^2\rangle \ge \langle\Psi(X)\rangle^2$. Note moreover that equality in \fer{d3} holds if and only if $\Psi(X)$ is constant, or, what is the same, if
 \[
  \frac{d^2}{dx^2}\log f(x) = c.
 \]
As observed in the proof of  Theorem \ref{bl} this implies that $X$ is a Gaussian variable.
\end{rem}

Grace to inequality \fer{d3}
 \be\label{d4}
 2I(f)I(g)\le 2\sqrt{J(f)J(g)} \le \frac ab J(f) + \frac ba J(g).
 \ee
Using \fer{d4} to bound from above the last term in inequality \fer{fin1} we obtain
 \be \label{new1}
 J(f*g) \le  \frac{a^3}{(a+b)^3} J(f) + \frac{b^3}{(a+b)^3} J(g).
 \ee
 Optimizing over $z = a/(a+b)$, with $z \in [0,1]$, one finds that the minimum of the right-hand side is obtained
 when
  \be\label{opt1}
 z = \bar z = \frac{\sqrt{J(g)}}{\sqrt{J(f)}+ \sqrt{J(g)}},
  \ee
  which implies
 inequality \fer{sharp2}. Thus we proved

\begin{cor}\label{cor-1}
Let $X$ and $Y$ be independent random variables with log-concave probability density
functions with values in $\R$,  such that both $J(X)$ and $J(Y)$, as given by \fer{j2-1} are bounded. Then
 \[
\frac 1{\sqrt{J(X+Y)}} \ge \frac 1{\sqrt{J(X)}}+ \frac 1{\sqrt{J(Y)}}.
 \]
Moreover, there is equality  if and only if,  up to translation and dilation  $X$ and
$Y$ are Gaussian variables.
\end{cor}

\begin{rem}
Inequality \fer{fin1} implies in general a stronger inequality. In fact, to obtain
inequality \fer{new1} we discarded the (non-positive) term
 \be\label{re5}
 -R(f,g,a,b) = \frac{a^2 b^2}{(a+b)^4}\left( 2I(f) I(g) -\frac ab J(f) - \frac ba J(g)\right).
 \ee
By evaluating the value of $R(f,g,a,b)$ in  $z = \bar z$,  one shows that inequality
\fer{fin1} is improved by the following
 \be\label{ex1}
 \frac 1{\sqrt{J(X+Y)}} \ge \left( \frac 1{\sqrt{J(X)}}+ \frac 1{\sqrt{J(Y)}}\right) \mathcal{R}(X,Y),
 \ee
 where
 \[
 1 \le \mathcal{R}(X,Y) = \left(1- 2\frac{\sqrt{J(X)}\sqrt{J(Y)}- I(X)I(Y)}{(\sqrt{J(X)}+\sqrt{J(Y)})^2} \right)^{-1/2}.
 \]
As before, $\mathcal{R}(X,Y)=1$ if and only if $X$ and $Y$ are Gaussian random
variables.

Note that the non-negative remainder $R(f,g,a,b)$ can be bounded from below in terms
of other expressions. In particular, one of these bounds is particularly
significative. Adding and subtracting to the right-hand side of \fer{re5} the positive
quantity $a I^2(f)/b +  b I^2(g)/a $ one obtains the bound
 \be\label{re6}
R(f,g,a,b) \ge \frac{a^2 b^2}{(a+b)^4}\left[ \frac ab( J(f)- I^2(f)) - \frac ba (J(g)- I^2(g))\right].
 \ee
This implies that \fer{new1} can be improved by the following
 \begin{equations} \label{new11}
 J(f*g) \le  \frac{a^3}{(a+b)^3} J(f) + \frac{b^3}{(a+b)^3} J(g) + \\ - \frac{a^2 b^2}{(a+b)^4}\left[ \frac ab( J(f)- I^2(f)) - \frac ba (J(g)- I^2(g))\right].
 \end{equations}
\end{rem}

\subsection{The general case}

With few variants, the proof in the multi-dimensional case follows along the same
lines of the one-dimensional one. Let $f(x)$ and $g(x)$, with $x \in \R^n$ be
multidimensional log-concave functions, and let $k(x) = f*g(x)$ be their log-concave
convolution. In addition, let us suppose that both $f(x)$ and $g(x)$ are sufficiently
smooth and decay at infinity in such a way to justify computations. To simplify
notations, given a function $f(x)$, with $x = (x_1, x_2, \dots, x_n) \in \R^n$, $n
>1$, we denote its partial derivatives as
 \[
 f_i(x)= \partial_if(x), \qquad f_{ij}(x) = \partial_{ij}f(x).
 \]
For any given vector $\alpha = (\alpha_1, \alpha_2, \dots , \alpha_n)$, and positive constants $a,b$ we have the identity
 \[
 (a+b) \sum_{i=1}^n \alpha_i \frac{k_i(x)}{k(x)} = \int_{\R^n}\sum_{i=1}^n\alpha_i \left(a \frac{f_i(x-y)}{f(x-y)} + b \frac{g_i(y)}{g(y)}\right) \,
d\mu_x(y),
 \]
where now, for every $x \in \R^n$
 \[
d\mu_x(y) = \frac{f(x-y)g(y)}{k(x)} \, dy,
 \]
is a unit measure on $\R^n$. Therefore, by Jensen's inequality
 \[
  (a+b)^2\left(\sum_{i=1}^n \alpha_i \frac{k_i(x)}{k(x)}\right)^2 = (a+b)^2 \sum_{i,j=1}^n\alpha_i\alpha_j \frac{k_i(x)}{k(x)} \frac{k_j(x)}{k(x)} \le
 \]
 \[
 \int_{\R^n}\sum_{i,j=1}^n\alpha_i\alpha_j \left( a^2 \frac{f_i(x-y)}{f(x-y)} \frac{f_j(x-y)}{f(x-y)} + b^2 \frac{g_i(y)}{g(y)} \frac{g_j(y)}{g(y)} + 2ab  \frac{f_i(x-y)}{f(x-y)}\frac{g_j(y)}{g(y)} \right)\, d\mu_x(y).
 \]
Likewise, thanks to the identity
\[
 (a+b)^2 \frac{k_{ij}(x)}{k(x)} = \int_{\R^n}\left(a^2 \frac{f_{ij}(x-y)}{f(x-y)} + b^2 \frac{g_{ij}(y)}{g(y)} + 2ab  \frac{f_{i}(x-y)}{f(x-y)}\frac{g_{j}(y)}{g(y)}\right) \,
d\mu_x(y),
 \]
we have
 \[
  (a+b)^2 \sum_{i,j=1}^n \alpha_i\alpha_j \frac{k_{ij}(x)}{k(x)} =
 \]
 \[
 \int_{\R^n}\sum_{i,j=1}^n\alpha_i\alpha_j \left(a^2 \frac{f_{ij}(x-y)}{f(x-y)} + b^2 \frac{g_{ij}(y)}{g(y)} + 2ab  \frac{f_{i}(x-y)}{f(x-y)}\frac{g_{j}(y)}{g(y)}\right) \,
d\mu_x(y).
 \]
Finally,  for any given vector $\alpha = (\alpha_1, \alpha_2, \dots , \alpha_n)$, and positive constants $a,b$ we obtain the inequality
 \begin{equations}\label{pd1}
  (a+b)^2 \sum_{i,j=1}^n \alpha_i\alpha_j \left( \frac{k_i(x)}{k(x)} \frac{k_j(x)}{k(x)} - \frac{k_{ij}(x)}{k(x)} \right) \le \\
 \sum_{i,j=1}^n \alpha_i\alpha_j \int_{\R^n} \left[ a^2 \left(\frac{f_i(x-y)}{f(x-y)} \frac{f_j(x-y)}{f(x-y)}- \frac{f_{ij}(x-y)}{f(x-y)} \right) \right. + \\
\left. b^2 \left(\frac{g_i(y)}{g(y)} \frac{g_j(y)}{g(y)} - \frac{g_{ij}(y)}{g(y)}  \right) \right] d\mu_x(y).
 \end{equations}
Inequality \fer{pd1} has important consequences. Indeed, the function $k(x)$, convolution of the two log-concave functions is log-concave. Consequently, the Hessian matrix $\mathcal{M}$ of $-(a+b)^2 \log k(x)$ is a symmetric positive semi-definite matrix, which implies that, for any given vector $\alpha = (\alpha_1, \alpha_2, \dots , \alpha_n)$
 \[
-  \sum_{i,j=1}^n \alpha_i\alpha_j(a+b)^2 (\log k(x))_{ij} = (a+b)^2 \sum_{i,j=1}^n \alpha_i\alpha_j \left( \frac{k_i(x)}{k(x)} \frac{k_j(x)}{k(x)} - \frac{k_{ij}(x)}{k(x)} \right) \ge 0.
 \]
 Likewise, since the functions $f$ and $g$ are log-concave, the matrix $\mathcal{N}$ with elements
  \[
  n_{ij} = \int_{\R^n} \left[ a^2 \left(\frac{f_i}{f} \frac{f_j}{f} - \frac{f_{ij}}{f}\right)(x-y) +
 b^2 \left(\frac{g_i}{g} \frac{g_j}{g} - \frac{g_{ij}}{g}  \right)(y) \right] d\mu_x(y)
  \]
is a symmetric positive semi-definite matrix.  Thus, inequality \fer{pd1} asserts that the matrix $\mathcal{N} - \mathcal{M}$ is itself a symmetric positive semi-definite matrix. Consequently, the product matrix $(\mathcal{N} - \mathcal{M})(\mathcal{N} + \mathcal{M})$ is positive semi-definite, which implies that its trace is non-negative. Now, in view of classical properties of the trace of a matrix we obtain
 \[
 {\tt{tr}} (\mathcal{N} - \mathcal{M})(\mathcal{N} + \mathcal{M}) = {\tt{tr}}\mathcal{N}\mathcal{N} + {\tt{tr}}\mathcal{N}\mathcal{M} - {\tt{tr}}\mathcal{M}\mathcal{N} - {\tt{tr}}\mathcal{M}\mathcal{M} =
 \]
 \[
{\tt{tr}}\mathcal{N}\mathcal{N} - {\tt{tr}}\mathcal{M}\mathcal{M} = \sum_{i,j=1}^n n_{ij}^2 - \sum_{i,j=1}^n m_{ij}^2 \ge 0.
 \]
Finally, inequality \fer{pd1} implies
 \begin{equations}\label{pd2}
   (a+b)^2 \sum_{i,j=1}^n \left( \frac{k_i(x)}{k(x)} \frac{k_j(x)}{k(x)} - \frac{k_{ij}(x)}{k(x)} \right)^2 \le \qquad\qquad\qquad \\
 \sum_{i,j=1}^n \left\{ \int_{\R^n} \left[ a^2 \left(\frac{f_i}{f} \frac{f_j}{f} - \frac{f_{ij}}{f}\right)(x-y) +
 b^2 \left(\frac{g_i}{g} \frac{g_j}{g} - \frac{g_{ij}}{g}  \right)(y) \right] d\mu_x(y)\right\}^2.
 \end{equations}
 Applying once more Jensen's inequality to the squares into the right-hand side of inequality \fer{pd2}, and then proceeding as in the proof of Theorem \ref{bl} we easily arrive to inequality \fer{ine-main}. Then, the cases of equality are found by the same argument of the one-dimensional proof. We conclude with the following

\begin{thm}\label{bn} Let $f(x)$ and $g(x)$  be log-concave probability density
functions with values in $\R^n$, with $n>1$,  such that both $J(f)$ and $J(g)$, as
given by \fer{j2} are bounded. Then, $J(f*g)$ is bounded, and for any pair of positive
constants $a,b$ \be\label{fin-n}
  J(f*g) \le  \frac{a^4}{(a+b)^4} J(f) + \frac{b^4}{(a+b)^4} J(g) + 2\frac{a^2 b^2}{(a+b)^4}H(f,g),
 \ee
where
 \[
 H(f,g) = \sum_{i,j=1}^n \int_{\{f>0\}} \frac{\partial_i f\partial_j f}f \, dx
 \int_{\{g>0\}} \frac{\partial_i g\partial_j g}g \, dx.
 \]
Moreover, there is equality in \fer{fin-n} if and only if, up to translation and dilation  $f$ and $g$ are Gaussian densities, $f(x)= M_{a}(x)$ and $g(x) = M_{b}(x)$.
\end{thm}

As for the one-dimensional case, given the random vector $X$ distributed with density $f(x)$, $x \in \R^n$, consider the generic element of the second-order score function $\mathcal{H}(X)$, given by \fer{scor2}
 \[
 \Psi_{ij}(X)  =  \frac{f_i(X)}{f(X)} \frac{f_j(X)}{f(X)} - \frac{f_{ij}(X)}{f(X)}.
 \]
Then, for each pair of $i,j$  it holds the identity
 \[
\langle \Psi_{ij}(X)\rangle = \int_{\{f>0\}} \frac{\partial_i f\partial_j f}f \, dx,\qquad  \langle   \Psi_{i,j}(X)^2\rangle = \int_{\{ f>0\}} \left[\frac{\partial_{ij}f}f -
\frac{\partial_{i}f\partial_{j}f}{f^2} \right]^2 f \, dx.
 \]
Then, the standard inequality $\langle \Psi_{ij}(X)^2\rangle \ge \langle\Psi_{ij}(X)\rangle^2$
gives
 \begin{equation}\label{mean1}
 J(X) = \sum_{i,j=1}^n \langle\Psi_{ij}(X)^2\rangle \ge \sum_{i,j=1}^n \langle\Psi_{ij}(X)\rangle^2 =  \sum_{i,j=1}^n \left[\int_{\{f>0\}} \frac{\partial_i f\partial_j f}f \, dx\right]^2
 \end{equation}
Using the Cauchy-Schwarz inequality, \fer{mean1} gives
\begin{equations}\label{bound4}
H(f,g) = \sum_{i,j=1}^n \int_{\{f>0\}} \frac{\partial_i f\partial_j f}f \, dx
 \int_{\{g>0\}} \frac{\partial_i g\partial_j g}g \, dx \le\qquad\qquad\qquad \\ \left\{\sum_{i,j=1}^n \left[\int_{\{f>0\}} \frac{\partial_i f\partial_j f}f \, dx\right]^2\right\}^{1/2}  \left\{\sum_{i,j=1}^n \left[\int_{\{g>0\}} \frac{\partial_i g\partial_j g}g \, dx\right]^2\right\}^{1/2} \le
\sqrt{J(f)}\sqrt{J(g)}.
\end{equations}
Hence, we can proceed as in the proof of Corollary \ref{cor-1} to obtain

\begin{cor}\label{cor-n}
Let $X$ and $Y$ be independent multi-dimensional random variables with log-concave probability density functions with values in $\R^n$,  such that both $J(X)$ and $J(Y)$, as given by \fer{j2} are bounded. Then
 \[
\frac 1{\sqrt{J(X+Y)}} \ge \frac 1{\sqrt{J(X)}}+ \frac 1{\sqrt{J(Y)}}.
 \]
Moreover, there is equality  if and only if,  up to translation and dilation  $X$ and $Y$ are Gaussian densities.
\end{cor}

\section{A strengthened entropy power inequality} \label{secEP}

In this section, we will study the evolution in time of the functional $\Lambda(t)$ defined in \fer{lh}, that is
 \[
 \Lambda(t)= \Phi(f(t),g(t)) = H(f(t)*g(t)) - \kappa H(f(t)) -(1-\kappa)H(g(t)).
 \]
Here,  $\kappa$ is a positive constant, with $0<\kappa<1$, while $f(x,t)$ (respectively
$g(x,t)$) are the solutions to the heat equation \fer{heat0} with diffusion constant
$\kappa$ (respectively $1-\kappa$), corresponding to the initial data $f$ and $g$, log-concave
probability densities in $\R^n$. It is a simple exercise to verify that $\Lambda(t)$ is dilation invariant. This property allows to identify the limit, as $t \to \infty$ of the functional $\Lambda(t)$.
For large times,  the solution to the heat equation approaches the fundamental
solution. This large-time behaviour can be better specified by saying that the
solution to the heat equation \fer{heat0} satisfies a property which can be defined as
the \emph{central limit property}. Suppose the initial density $f$ in equation \fer{heat0} is such
that $(1 + |x|^\delta)f(x) \in L^{1}(\R^n)$ for some constant $\delta >0$ (typically
$\delta =2$). Then, if
\begin{equation}\label{FPscal}
 U(x,t) = \left(\sqrt{1+2t}\right)^n\, u(x\, \sqrt{1+2t}, t).
\end{equation}
$U(x,t)$ tends in $L^{1}(\R^n)$ towards a limit function as time goes to infinity, and
this limit function is a Gaussian function
 \be\label{limi8}
\lim_{t\to \infty} U(x,t) = M_\kappa(x) \, \int_{\R^n} f(x) \, dx =
 M_\kappa(x).
 \ee
This convergence property, as well as convergence in other stronger norms, can be
achieved easily by resorting to Fourier transform, or by exploiting the relationship
between the heat equation and the Fokker--Planck equation \cite{CT, GJT} (cf. also
\cite{BBDE} for recent results and references). We note that the passage $u(x,t) \to
U(x,t)$ defined by \fer{FPscal} is dilation invariant, so that
 \be\label{mp}
\int_{\R^n} U(x,t) \, dx =   \int_{\R^n} u(x,t) \, dx.
 \ee
Coupling the dilation invariance of $\Lambda(t)$ with the central limit property, and remarking that $M_{\kappa}*M_{1-\kappa}= M_1$, gives
 \begin{equations}\label{inf1}
 \lim_{t\to \infty}\Lambda(t) = H(M_1) - \kappa H(M_\kappa) - (1-\kappa)H(M_{1-\kappa}) = \\
 -\frac n2 \left[ \kappa \log \kappa +
(1-\kappa)\log(1-\kappa)\right].\qquad\qquad\quad
 \end{equations}
If we differentiate $\Lambda(t)$ with respect to time, by the Bruijn's identity we obtain
 \be\label{fis3}
\Lambda'(t)= I(f(t)*g(t)) - \kappa^2 I(f(t)) -(1-\kappa)^2I(g(t)).
 \ee
In view of the inequality \cite{Bla, Sta}
 \[
I(f*g) \le \frac{a^2}{(a+b)^2} I(f) + \frac{b^2}{(a+b)^2}I(g), \quad a,b >0
 \]
with equality if and only if $f$ and $g$ are Gaussian densities, $\Lambda'(t) \le 0$.
Hence $\Lambda(t)$ is monotonically decreasing in time from $\Lambda (0)$ to
$\Lambda(\infty)$, given by \fer{inf1}. Moreover, optimizing over $\kappa$, one shows
that the non-negative quantity
 \[
 H(f*g) - \kappa H(f) -(1-\kappa)H(g) -\frac n2 \left[ \kappa \log \kappa +
(1-\kappa)\log(1-\kappa)\right]
 \]
attains the maximum
when
 \be\label{max1}\displaystyle
 \kappa = \bar \kappa = \frac{\exp \left\{ \frac 2n H(f)  \right\}}{\exp \left\{ \frac 2n H(f)  \right\}+ \exp \left\{ \frac 2n H(g)  \right\} },
 \ee
Evaluating $\Lambda(0) - \Lambda(\infty)$ in $\kappa= \bar\kappa$ gives the entropy power inequality \fer{entr}. Note that this result holds for all probability densities.

Differentiating again with respect to time, from \fer{fis3} we obtain
 \be\label{fis4}
\Lambda''(t)= -\frac 12\left( J(f(t)*g(t)) - \kappa^3 J(f(t)) -(1-\kappa)^3J(g(t))\right).
 \ee
Therefore, by inequality \fer{new1}, if $f$ and $g$ are log-concave, $\Lambda''(t) \ge0$, and the convexity property of $\Lambda(t)$ follows.

On the other hand, proceeding as in the proof of Corollary \ref{cor-1},  we obtain from  inequality \fer{fin-n} the bound
 \be\label{resto}
 J(f(t)*g(t)) \le \kappa^3 J(f(t)) +(1-\kappa)^3J(g(t)) - 2\kappa^2(1-\kappa)^2 P(f(t), g(t)),
 \ee
where
\be\label{resto1}
 P(f, g) = \sqrt{J(f)}\sqrt{J(g)} -H(f,g) \ge0
\ee
in view of inequality \fer{bound4}. In addition, equality to zero holds if and only if both $f$ and $g$ are Gaussian densities.

Integrating \fer{resto} from $t$ to $\infty$, we obtain for the Fisher information of two log-concave densities the strengthened inequality
 \be\label{str1}
  I(f(t)*g(t)) \le \kappa^2 I(f(t)) -(1-\kappa)^2I(g(t)) - \kappa^2(1-\kappa)^2 \int_t^\infty P(f(s), g(s))\, ds.
 \ee
In fact, by the central limit property,
 \[
 \lim_{t\to\infty} \left[ I(f(t)*g(t)) - \kappa^2 I(f(t)) -(1-\kappa)^2I(g(t)) \right]=
  \]
  \[
  I(M_1) -  \kappa^2 I(M_\kappa) -(1-\kappa)^2I(M_{1-\kappa}) = 0
 \]
Last, integrating \fer{str1} from $0$ to $\infty$ we obtain for Shannon's entropy of
the two log-concave densities the strengthened inequality
 \be\label{str2}
H(f*g) - \kappa H(f) -(1-\kappa)H(g) -\frac n2 \left[ \kappa \log \kappa +
(1-\kappa)\log(1-\kappa)\right] \ge \mathcal{P}_\kappa(f,g),
 \ee
where
\be\label{res3}
 \mathcal{P}_\kappa(f,g) =  \kappa^2(1-\kappa)^2 \int_0^\infty d\tau \int_\tau^\infty P(f(s), g(s))\, ds \ge 0.
 \ee
Choosing now $\kappa = \bar\kappa$ as given by \fer{max1} we end up with  inequality \fer{sep}, where
 \be\label{ma1}
 R(X,Y) = \exp\left\{ \frac 2n\mathcal{P}_{\bar\kappa}(f,g)\right\} >1.
 \ee
Note that the term $R(X,Y)$ is related to the second-order score of the random vectors $X$ and $Y$. Consequently, $R(X,Y) = 1$ if and only if both $X$ and $Y$ are Gaussian random vectors.

\begin{rem}
In general, the expression of the term $R(X,Y)$ is very complicated, due to the fact
that it is given in terms of integrals of nonlinear functionals evaluated along
solutions to the heat equations which depart from the densities of $X$ and $Y$. It
would be certainly interesting to be able to express the term $R(X,Y)$ (or to bound it
from below)  in terms of some distance of $X$ and $Y$ from the space of Gaussian
vectors. This problem is clearly easier in one dimension, where one can use the
remainder as given by inequality \fer{new11}, namely as the sum of the two
contributions of the type $J(f) - I^2(f)$. In this case, one would know if, for some
distance $d(f,g)$ between two probability densities $f$ and $g$ and some positive
constant $C$
 \[
 J(f) - I^2(f) \ge C \inf_{M \in \mathcal{M}} d(f, M),
 \]
 where $\mathcal{M}$ denotes the space of Gaussian densities.
\end{rem}

\section{Conclusions}

In this paper, we analyzed various inequalities for convolutions for log-concave
densities. The main discovery is that log-concave densities satisfy a new inequality
for convolutions which appears as the natural generalization of Shannon's entropy
power \fer{entr} and  Blachman--Stam  \fer{BS} inequalities. This inequality is sharp,
and it is the starting point for deriving Shannon's entropy power and Blachman--Stam
inequalities in a strengthened form. It results in a clear way from the present
analysis, that the behavior of the log-concave density functions with respect to
convolutions deserves further investigations.

\bigskip \noindent

{\bf Acknowledgment:} This work has been written within the activities of the National Group of Mathematical Physics of INDAM (National Institute of High Mathematics). The  support of the  project ``Optimal mass
transportation, geometrical and functional inequalities with applications'', financed by the Minister of University and Research, is kindly acknowledged.

\end{document}